\documentclass[pra,twocolumn,aps,superscriptaddress,showpacs,floatfix]{revtex4-2}

\usepackage{amsmath}
\usepackage{amssymb}
\usepackage{amsbsy}
\usepackage{graphicx}
\usepackage{color}
\usepackage{epstopdf}
\usepackage{mathrsfs}

\begin{document}
	
	\title{Emergence of multifractality through cascade-like transitions in a mosaic interpolating Aubry-Andr\'{e}-Fibonacci chain}
	\author{Qi Dai}
	\affiliation{Institute of Theoretical Physics and State Key Laboratory of Quantum Optics and Quantum Optics Devices, Shanxi University, Taiyuan 030006, China}
	\author{Zhanpeng Lu}
	\affiliation{Institute of Theoretical Physics and State Key Laboratory of Quantum Optics and Quantum Optics Devices, Shanxi University, Taiyuan 030006, China}	
	\author{Zhihao Xu}
	\email{xuzhihao@sxu.edu.cn}
	\affiliation{Institute of Theoretical Physics and State Key Laboratory of Quantum Optics and Quantum Optics Devices, Shanxi University, Taiyuan 030006, China}
	\affiliation{Collaborative Innovation Center of Extreme Optics, Shanxi University, Taiyuan 030006, China}

	\begin{abstract}
		In this paper, we explore the localization features of wave functions in a family of mosaic quasiperiodic chains obtained by continuously interpolating between two limits: the mosaic Aubry-Andr\'{e} (AA) model, known for its exact mobility edges with extended states in the band-center region, and localized ones in the band-edge regions for a large enough modulation amplitude, and the mosaic Fibonacci chain, which exhibits its multifractal nature for all the states except for the extended one with $E=0$ for an arbitrary finite modulation amplitude. We discover that the mosaic AA limit for the states in the band-edge regions evolves into multifractal ones through a cascade of delocalization transitions. This cascade shows lobes of lower fractal dimension values separated by maxima of fractal dimension. In contrast, the states in the band-center region (except for the $E=0$ state) display an anomalous cascading process, where it emerges lobes of higher fractal dimension values are separated by the regions with lower fractal dimensions. Our findings offer insight into understanding the multifractality of quasiperiodic chains.
	\end{abstract}

	
	\maketitle
	\section{Introduction}
	Quasicrystals, as one of the essential solid structures, constitute a distinctive phase between fully periodic lattices and fully disordered media, exhibiting a structure that is long-range ordered but not periodic. Quasiperiodic systems \cite{Lellouch,Jazaeri,Yuce,Q,Longhi,Jiang} demonstrate exotic conduction features, self-similar structures, and critical behaviors. The experimental developments in cold atoms \cite{Roati,Delande,Modugno,Roux,D'Errico,Giamarchi,Sbroscia,Gautier,Yu} and photonic crystals \cite{Oton,Lahini} have made the study of the dynamics of quasicrystals both in one-dimensional (1D) and two-dimensional quasiperiodic systems experimentally accessible. These impressive quasiperiodic structures have drawn great focus, including their nontrivial connection to topological phases \cite{Kraus} and a variety of localization transitions between extended, localized, and critical phases \cite{Schiffer,Yao,Kalugin,Aubry,Han,Hiramoto,Kohmoto}.
	
    Many theoretical quasiperiodic models \cite{Lahini,Schiffer,Yao,Kalugin,Aubry,Han,Hiramoto,Kohmoto,Jitomirskaya,Xia} have been proposed to study their localization transitions and the critical phenomena. Due to the simplicity and experimental realization \cite{Kalugin,Aubry,Jitomirskaya,Zeng,Yang,Xu,Sinha,Li,Mastropietro,Ganeshan,Harper,Jagannathan,Kohmoto,Xia,Ostlund,Merlin,E,Ashraff,F,Verbin,Y,Goblot}, the Anbry-Andr\'{e} (AA) model \cite{Aubry,Jitomirskaya,Xia,Zeng,Yang,Xu,Sinha,Li,Mastropietro,Ganeshan,Harper} and the Fibonacci model \cite{Jagannathan,Ostlund,Merlin,E,Ashraff,F} are two paradigmatic examples of 1D quasicrystal systems that have been widely studied. The specific properties of the AA model with an on-site incommensurate modulation is that above a finite critical modulation amplitude, all the eigenstates change from extended to localized, determined by the unique self-dual characteristic \cite{Aubry,Jitomirskaya}. In contrast, the modulation of the Fibonacci chain exhibits two discrete values that appear interchangeably according to the Fibonacci sequence. The Fibonacci model always has critical wave functions for an arbitrary value of the on-site modulation. Though two such paradigmatic models have very different localization features, they have many connections, such as they belong to the same topological class and are two limits of the interpolating Aubry-Andr\'{e}-Fibonacci (IAAF) model \cite{Verbin,Jagannathan,Kraus}, which is proposed to discuss the topological relationship and how criticality develops during a smooth interpolation between the AA model and the Fibonacci chain. Recently, Goblot {\it et al.} \cite{Goblot} theoretically and experimentally demonstrated the occurrence of a cascade of delocalization transition as the model is tuned from the AA to the Fibonacci limit. Moreover, one finds that the nonreciprocal hopping can drastically change the cascade behavior in the non-Hermitian IAAF model \cite{Zhai}. Some quasiperiodic models with long-range hopping also display such cascading phenomena \cite{Roy,Zhi}.
	
	On the other hand, the mobility edge as a crucial concept in disordered systems, which separates extended and localized single-particle states in the energy spectra, may lead to some fundamental physics \cite{Evers}, such as the metal-insulator transition and the thermoelectric response \cite{Whitney,Chiaracane,Yamamoto}. Great efforts have been made in exploring quasiperiodic systems with mobility edges. By introducing an energy-dependent self-duality, one can obtain some generalized AA models with exact mobility edges, such as 1D quasicrystals with long-range hopping \cite{Biddle,Priour,Xiao,Shina,X,D,Liu} or a unique form of the on-site incommensurate modulation \cite{Pixley,Alex,He,Xie}. Recently, by using mathematical tools, a class of more generic models with mobility edges, which can be exactly solved beyond the dual transformation, has been introduced. It is highly significant to explore the rich mobility edge physics further. Specially, through computing the Lyapunov exponents \cite{Shu,Zhang} from Avila's global theory \cite{Avila,A,You}, Wang {\it et al.} exhibited a 1D quasiperiodic mosaic chain with exact mobility edges \cite{Wang,Zhou}, which can be realized for ultracold atoms based on an optical Raman lattice \cite{Roati}.
	
	Here we study the fate of the cascadelike transitions in different band regions for a mosaic IAAF model. The extended states emerge in the band-center region in the mosaic AA limit for an arbitrary quasiperiodic modulation amplitude $\lambda$, and in the strong disorder regime, the states in the band-edges regions are localized. The numerical and analytical results show that the system exhibits exact mobility edges. In contrast, all the eigenstates in the mosaic Fibonacci limit are multifractal except for the state with $E=0$. Unlike the former IAAF case, the mosaic IAAF model displays two different ways to enter the multifractal phase along the continuous deformation from the mosaic AA limit to the mosaic Fibonacci limit; that is, the cascadelike transitions display different behaviors for the states in different band regions. In the band-edge regions, the cascade of the fractal dimension is found, similar to that found in the standard IAAF model. However, with the increase of the knob parameter in the band-center region, we exhibit an anomalous cascaded process with the emergence of the lower fractal dimension values between the regions with higher fractal dimension values.

	The plan of this paper is as follows. In Sec. II, the mosaic IAAF model is presented, and the order parameters to characterize three different types of states are listed. In Sec. III, we discuss the localization features and critical behavior of the mosaic AA and Fibonacci limits. We continuously control the knob parameter and present the cascadelike transitions for the states in different band regions shown in Sec. IV. A conclusion and the experimental possibilities of the detection of the cascading transitions are given in Sec. V.
	
	\section{Mosaic IAAF model and methods}
	We consider a mosaic IAAF model is a 1D tight-binding chain with a quasiperiodic mosaic on-site modulation, which can be described by	
	\begin{equation}\label{eq1}
		\hat{H} = t \sum_{j} (\hat{c}_j^{\dagger}\hat{c}_{j+1} + \rm{H.c.}) + \lambda \sum_{j} V_j(\beta) \hat{c}_j^{\dagger}\hat{c}_j,
	\end{equation}
	where $\hat{c}_j$ is the annihilation operators at site $j$; $t$ is the nearest-neighbor hopping amplitude, which is set as an energy unit, i.e., $t=1$; and $\lambda$ is the strength of the on-site quasiperiodic modulation. The on-site potential $V_j(\beta)$ is defined as
	\begin{equation}\label{eq2}
		V_j(\beta) = \begin{cases}  - \frac{\tanh[\beta(\cos{(2\pi\alpha m)}-\cos{(\pi \alpha)})]}{\tanh{\beta}},  & j = 2m, \\
			0,  & j=2m-1, \\
		\end{cases}
	\end{equation}
	with the tunable parameter $\beta$, $m$ being the index of quasicells, and the spatial modulation frequency set as the inverse of the golden mean, $\alpha=(\sqrt{5}-1)/2$. Since the quasiperiodic potential periodically occurs at the even sites, and the modulation amplitude of the odd sites is set to zero, we can introduce a quasicell with the nearest two lattice sites. If the number of the quasicell is $N$, i.e., $m=1,2,\dots,N$, the size of the system will be $L=2N$. The tunable parameter $\beta$ provides a knob by which we can interpolate between two limiting cases: (1) For the $\beta \to 0$ limit, the system is described by a 1D quasiperiodic mosaic lattice with the potential at the even sites $V_j(\beta)=\cos{(2\pi \alpha m)} - \cos{(\pi \alpha)}$, and the odd sites' amplitude is zero. (2) For the $\beta \to\infty$ limit, $V_{j}(\beta)$ corresponds to a step potential switching between $\pm 1$ values according to the Fibonacci sequence at the even sites \cite{Kohmoto,Ostlund}, and the potential amplitude of the odd sites is equal to zero. Supposing that the eigenstate of the mosaic IAAF chain is given by $|\psi_j\rangle = \sum_{j} \psi_j \hat{c}_j^{\dagger}|0\rangle$, the eigenvalues of the system can be obtained from the characteristic equations:
	\begin{equation}
		\begin{cases}
			\psi_{2m+1}+\psi_{2m-1}+\lambda V_{2m}\psi_{2m} = E \psi_{2m}, \notag \\
			\psi_{2m+2}+\psi_{2m} = E \psi_{2m+1}, \notag		
		\end{cases}
	\end{equation}
	where $\psi_{j}$ is the amplitude of the eigenstate at $j$th site and $E$ is the eigenvalue. One can easily obtain the reduced eigenvalue equation for $E\ne 0$ as follows:
	\begin{equation}\label{eq3}
		\psi_{2m+2}+\psi_{2m-2}+\lambda E V_{2m}\psi_{2m} = E \left(E-\frac{2}{E}\right) \psi_{2m}.
	\end{equation}
	For the $E=0$ case, one can apply the transfer matrix of the Schr\"{o}dinger operator in one quasicell $T_m(E=0)$ to obtain the corresponding Lyapunov exponent $\mathcal{L}(0)=\lim_{N\to \infty} \ln{||T(0)||}/(2N)$, where $||A||$ denotes the norm of the matrix $A$, and the total transfer matrix $T(0)=T_N(0)T_{N-1}(0)\cdots T_1(0)$. When the Lyapunov exponent $\mathcal{L}\to 0$, the corresponding state is a delocalized one, and for a finite Lyapunov exponent, it is a localized state. The transfer matrix for one quasicell with $E=0$ is an upper triangular matrix $T_m(0)=\begin{pmatrix} -1 & \lambda V_{2m} \\ 0 & -1 \end{pmatrix}$. Hence, the total transfer matrix  $T(0)=(-1)^N \begin{pmatrix} 1 & -\lambda (V_2+V_4+\cdots +V_{2N}) \\ 0 & 1 \end{pmatrix}$, and the norm of the total transfer matrix $||T(0)||=1$, which corresponds to an extended state with $\mathcal{L}(0)=0$ independent of $\beta$ and $\lambda$.
	
	To characterize the localization property of the wave function, one can calculate the inverse participation ratio (IPR) \cite{Evers},
	\begin{equation}\label{eq4}
		\mathrm{IPR}^{(n)} = \sum_{j}|\psi_j^{(n)}|^4,
	\end{equation}
	for the $n$th eigenstate with the eigenvalue $E_n$. In the region where the eigenstate $\psi_n$ is extended, the IPR is equal to the inverse of the chain length and tends to zero in the thermodynamic limit. For a localized state, the IPR remains a finite value with the increase of the system size. For a multifractal state, $\mathrm{IPR}^{(n)} \propto L^{-\eta_n}$ with $\eta_n \in (0,1)$. Hence, one can define the fractal dimension of the $n$th eigenstate $\psi_j^{(n)}$ as follows:
	\begin{equation}\label{eq5}
		\eta_n = -\lim_{L\to\infty} \left[\frac{\ln\mathrm{IPR}^{(n)}}{\ln{L}}\right].
	\end{equation}
	According to the above discussion, it is known that $\eta_n \to 1$ for an extended state; $\eta_n \to 0$ for a localized state; and when $0<\eta_n<1$, the state corresponds to a multifractal one. To avoid the fluctuation of a fixed state at different sizes, it is convenient to study the mean IPR (MIPR) $\mathrm{MIPR}=(1/L)\sum_n \mathrm{IPR}^{(n)}$.
	
	The detection of three different types of states also can be achieved by contrasting the distribution of the states in real and momentum spaces. The distribution of the states in the momentum space \cite{Sun} is given by
	\begin{equation}\label{eq6}
		n_k = \langle \psi|\hat{c}_{k}^{\dagger}\hat{c}_k |\psi \rangle,
	\end{equation}
	where $\hat{c}_k = (1/\sqrt{L}) \sum_{j} e^{ikj} \hat{c}_j$. In the momentum space, the extended (localized) state displays the localized (extended) distribution [see Figs.  \ref{Fig2}(a)$-$2(d)]. For a multifractal state, the distribution exhibits a delocalized and nonergodic behavior in both spaces [see Figs. \ref{Fig4}(a)$-$4(d)]. Similar to the fractal dimension the real space, we define the fractal dimension in the momentum space,
	\begin{equation}\label{eq7}
		\eta_n^{(k)} =-\lim_{L\to\infty} \left[\frac{\ln\mathrm{IPR}^{(n)}_k}{\ln{L}}\right],
	\end{equation}
	where the IPR in the momentum space $\mathrm{IPR}_k^{(n)} = \sum_{l} n_{k_l}^2$ with $k_l=2\pi l/L$ ($l=0,1,\dots,L-1$). For extended (localized) states in the real space, $\eta_n^{(k)}$ extrapolate to $0$ ($1$), while the values of $\eta_n^{(k)}$ are far from $0$ and $1$ in the multifractal zone.
	
	In this paper, the parameter of the modulation $\alpha$ can be approximately obtained by considering a Fibonacci sequence \cite{Kohmoto,Ostlund}, $F_{v+1} = F_{v} + F_{v-1}$, with $F_0=F_1=1$. We take the system size $L=2N=2F_v$ and the rational approximation $\alpha=F_{v-1}/F_v$. We apply exact diagonalization method to numerically study the mosaic IAAF model Eq. (\ref{eq1}) under periodic boundary conditions (PBCs). Moreover, the eigenvalues are ordered in ascending order.
	
	\section{Localization features in mosaic AA and Fibonacci limits}
	
	\subsection{The mosaic AA limit}
	
	\begin{figure}[tbp]
		\begin{center}
			\includegraphics[width=.5 \textwidth] {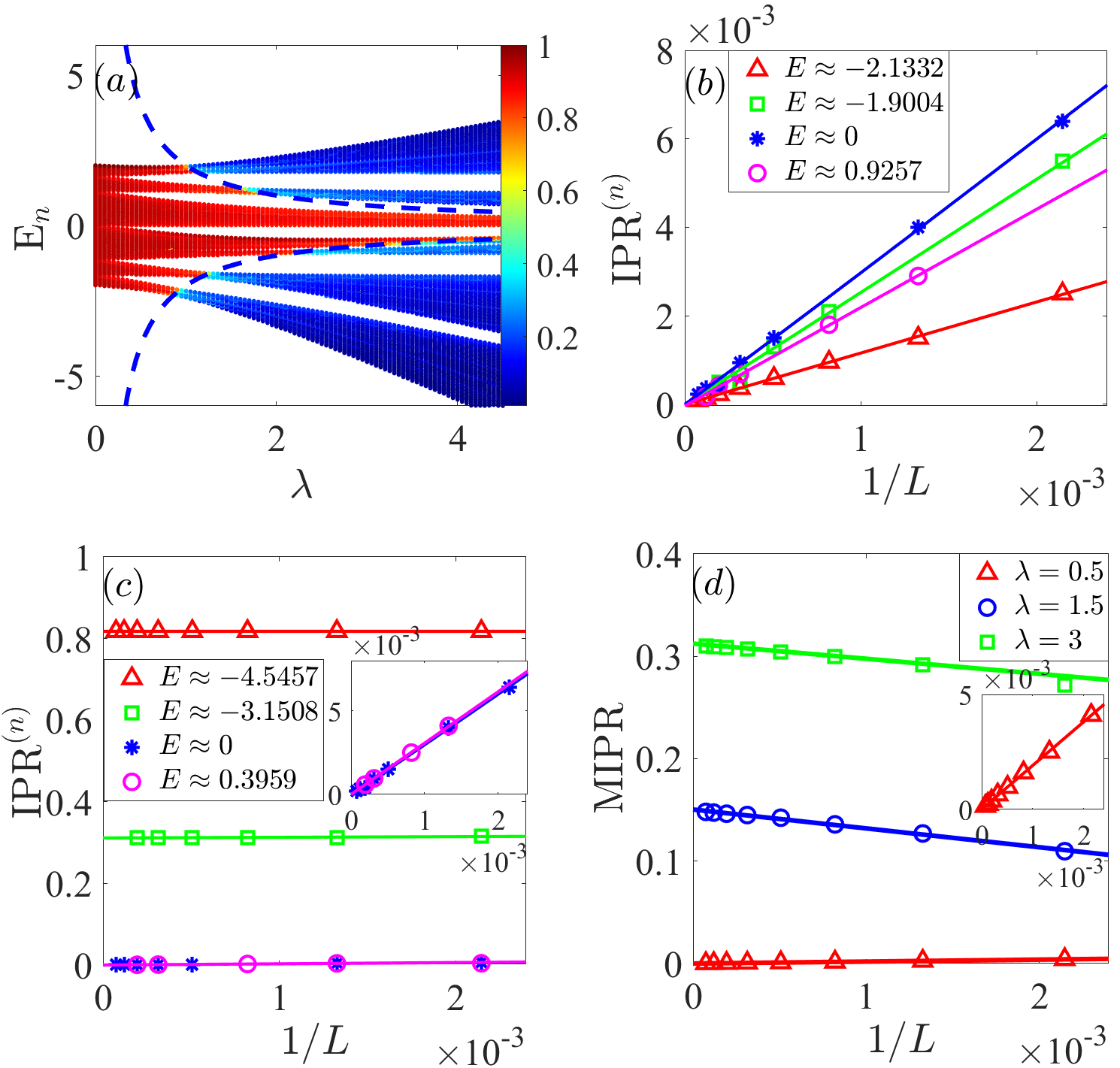}
		\end{center}
		\caption{(Color online) (a) Fractal dimension $\eta_n$ of different eigenstates as a function of the eigenvalues $E_n$ and the quasiperiodic potential amplitude $\lambda$ with $L=2N=1220$. The blue dashed lines represent the mobility edges given by $E_c=\pm 2/\lambda$. The scaling of $\rm{IPR}s$ for different eigenstates with (b) $\lambda =0.5$ and (c) $\lambda =3$, respectively. The inset in (c) shows the enlarge view of the scaling of IPRs for the eigenstates with $E=0$ and $0.3959$ for $\lambda=3$, respectively.(d) The scaling of MIPRs with different $\lambda$. The inset in (d) shows the enlarge view of the $\lambda=0.5$ case. Here, $\beta\to 0$.}\label{Fig1}
	\end{figure}
	
	The $\beta \to 0$ limit corresponds to the mosaic AA model for Eq. (\ref{eq1}). The reduced characteristic equation Eq. (\ref{eq3}) is similar to the AA model in this limit. For an AA model, the metal-insulator transition emerges at the modulation amplitude above two times the hopping amplitude. Thus, we can analytically obtain the mobility edges of the quasiperiodic mosaic lattice $E_c=\pm 2/\lambda$ \cite{Wang,Xia}. A fundamental feature of such a model is that for an arbitrarily strong quasiperiodic potential, the mobility edges always take place [see Fig. \ref{Fig1}(a)], which is the energy separating the extended and the localized states of the system. Figure \ref{Fig1}(a) shows the fractal dimension $\eta_n$ of different eigenstates in the real space as a function of the corresponding energies and the modulation amplitude $\lambda$ for the mosaic IAAF model in the $\beta \to 0$ limit. As shown in Fig. \ref{Fig1}(a), the analytical results of the mobility edges are marked by the blue dashed lines, and the energies of the extended states always emerge in the band-center region of the spectrum, in which the fractal dimensions approach unit. When $\lambda<\lambda_{c}\approx 0.86$, all the states of the system are extended and, for $\lambda>\lambda_{c}$, the mobility edges emerge. In Figs. \ref{Fig1}(b) and 1(c), we show the scaling of IPRs in the $\beta=0$ limit for different eigenstates with $\lambda=0.5$ and $3$, respectively. For $\lambda=0.5$ ($<\lambda_c$), when $L\to \infty$, the IPRs of different states approach $0$ with $\eta_n \to 1$. For $\lambda=3$, the $E=0$ and $E \approx 0.3959$ eigenstates localized in  $[-2/\lambda,2/\lambda]$ correspond to the extended state with $\eta_n \to 1$ [see the inset of Fig. \ref{Fig1}(c)], and when the eigenvalues of the states below (above) $-2/\lambda$ ($2/\lambda$), the IPRs of such states are independent of the system size, with $\eta_n \to 0$ corresponding to the localized states. As shown in Fig. \ref{Fig1}(c) for $\lambda=3$, the band-edge states with $E\approx -4.5457$ and $E\approx -3.1508$, both which are below $-2/\lambda$, exhibit localization properties. Figure \ref{Fig1}(d) shows the MIPR as a function of $1/L$ for different $\lambda$. In the fully extended regime ($\lambda=0.5$), the MIPR approaches $1/L$ with the increase of system size and drops to $0$ in the infinite size limit. When the system enters the regime with mobility edges, $\mathrm{MIPR}$ tends to a finite value in the thermodynamic limit. As seen in Fig. \ref{Fig1}(d), the stronger the quasiperiodic modulation amplitude $\lambda$, the larger the value of $\mathrm{MIPR}$ in the thermodynamic limit.
	
\begin{figure}[tbp]
	\begin{center}
		\includegraphics[width=.5 \textwidth] {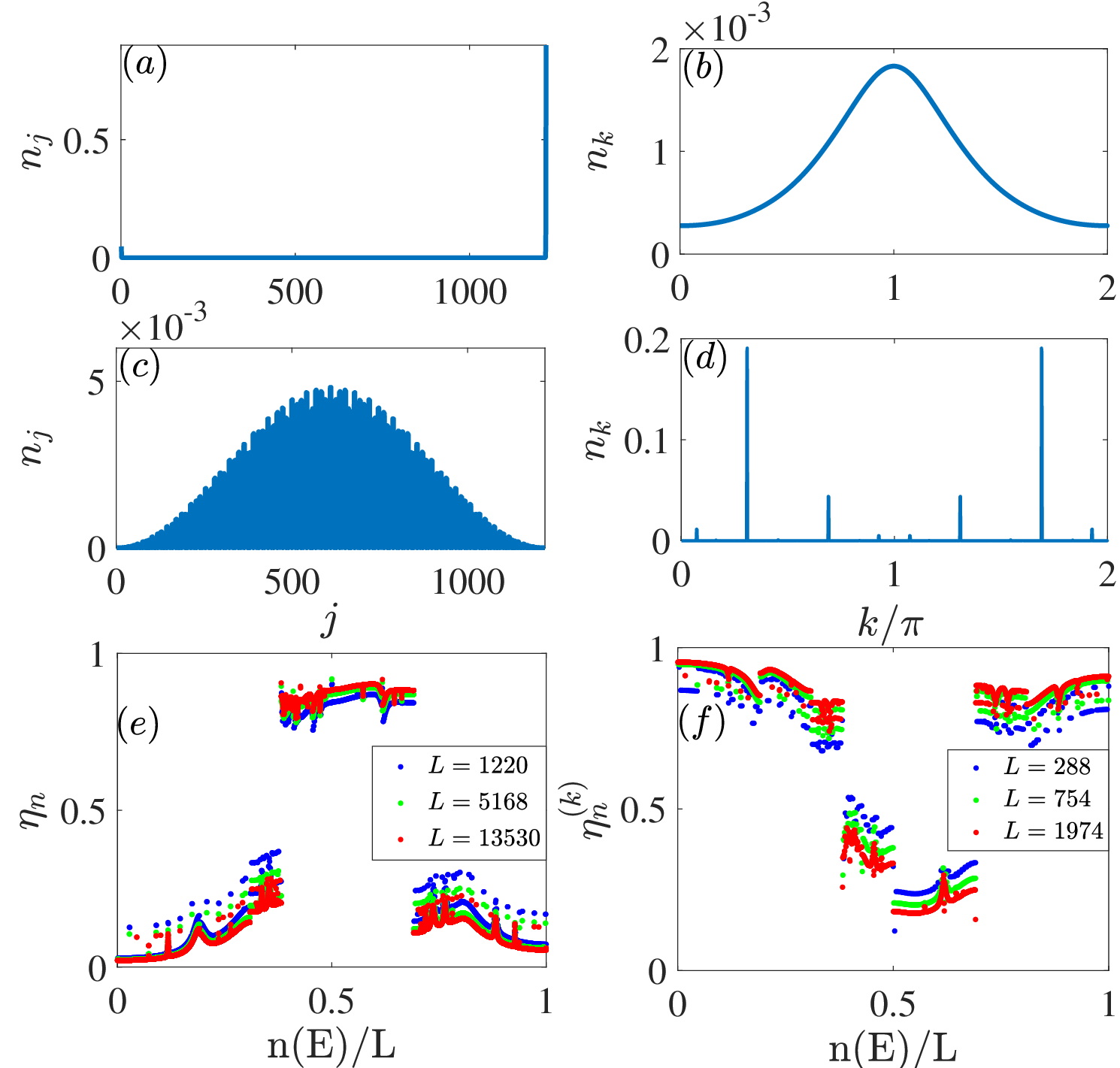}
	\end{center}
	\caption{(Color online) (a), (b) Distribution of the first with $E\approx -4.5457$ eigenstate in the real and momentum spaces, respectively. (c), (d) Distribution of the $843$th with $E\approx 0.3959$ eigenstate in the real and momentum, respectively. (e), (f) Fractal dimensions in the real and momentum spaces for different sizes, respectively. Here, $L=2N=1220$ for [(a)$-$(d)], $\lambda=3$ and $\beta  \to 0$.}\label{Fig2}
\end{figure}

	Figures \ref{Fig2}(a)$-$\ref{Fig2}(d) show the distributions of different eigenstates in the real and momentum spaces, respectively. The distribution $n_j=|\psi_j|^2$ of the first with $E\approx -4.5457$ ($843$th with $E\approx 0.3959$) eigenstate for $\lambda=3$, $L=2N=1220$, and $\beta=0$ in the real space exhibits localized (extended) features shown in Fig. \ref{Fig2}(a) [Fig. \ref{Fig2}(c)], while in the momentum space, the corresponding distribution is extended (localized), which is shown in Fig. \ref{Fig2}(b) [Fig. \ref{Fig2}(d)]. By contrasting the fractal dimensions for each eigenstate at different system sizes in real and momentum spaces, one can obtain clear information on the localization properties of the system in the mosaic AA limit, shown in Figs. \ref{Fig2}(e) and \ref{Fig2}(f) with $\lambda=3$, and $\beta=0$, respectively. In the finite-size case, the fractal dimension of the states in the localized regions extrapolates to $0$ and $1$ in real and momentum spaces, respectively. In contrast, the extended states' fractal dimensions tend to be $1$ and $0$ with increased system size in both spaces.
	
	The analytical and numerical results indicate that the system has exact mobility edges in the mosaic AA limit, and the extended states emerge in the band-center region for an arbitrary modulation amplitude. In a large $\lambda$ case, the localized states emerge in the band-edge region. With the increase of $\lambda$, the extended regime localized in the band-center region shrinks.
	
\subsection{The mosaic Fibonacci limit}

\begin{figure}[tbp]
	\begin{center}
		\includegraphics[width=.5 \textwidth] {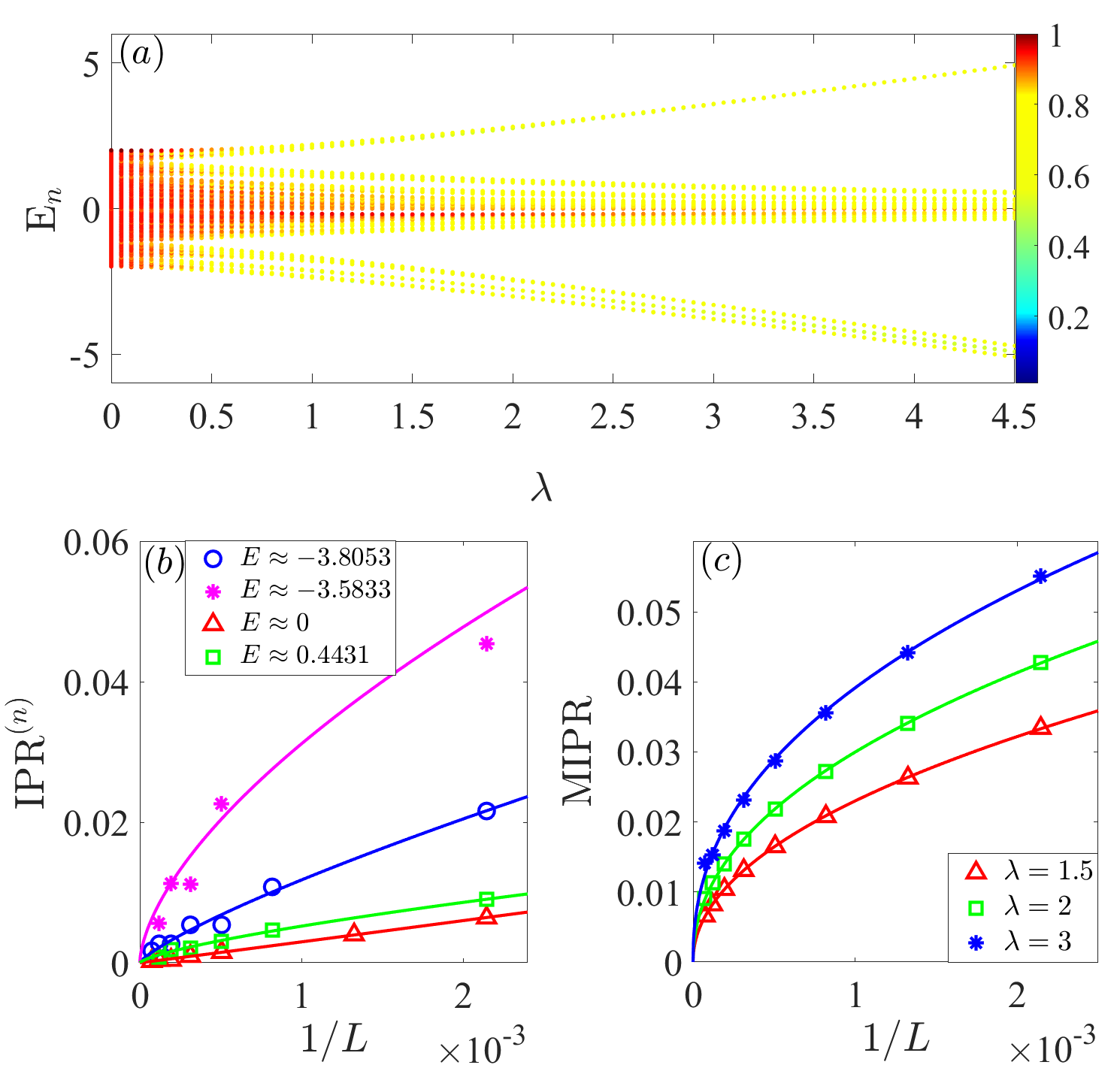}
	\end{center}
	\caption{(Color online) (a) Fractal dimension $\eta_n$ of different eigenstates as a function of the eigenvalues $E_n$ and the quasiperiodic potential amplitude $\lambda$ with $L=2N=1220$. (b) The scaling of $\rm{IPRs}$ for different eigenstates with $\lambda =3$. (c) The scaling of MIPRs for different $\lambda$. Here, $\beta  \to \infty$.}\label{Fig3}
\end{figure}

	\begin{figure}[tbp]
	\begin{center}
		\includegraphics[width=.5 \textwidth] {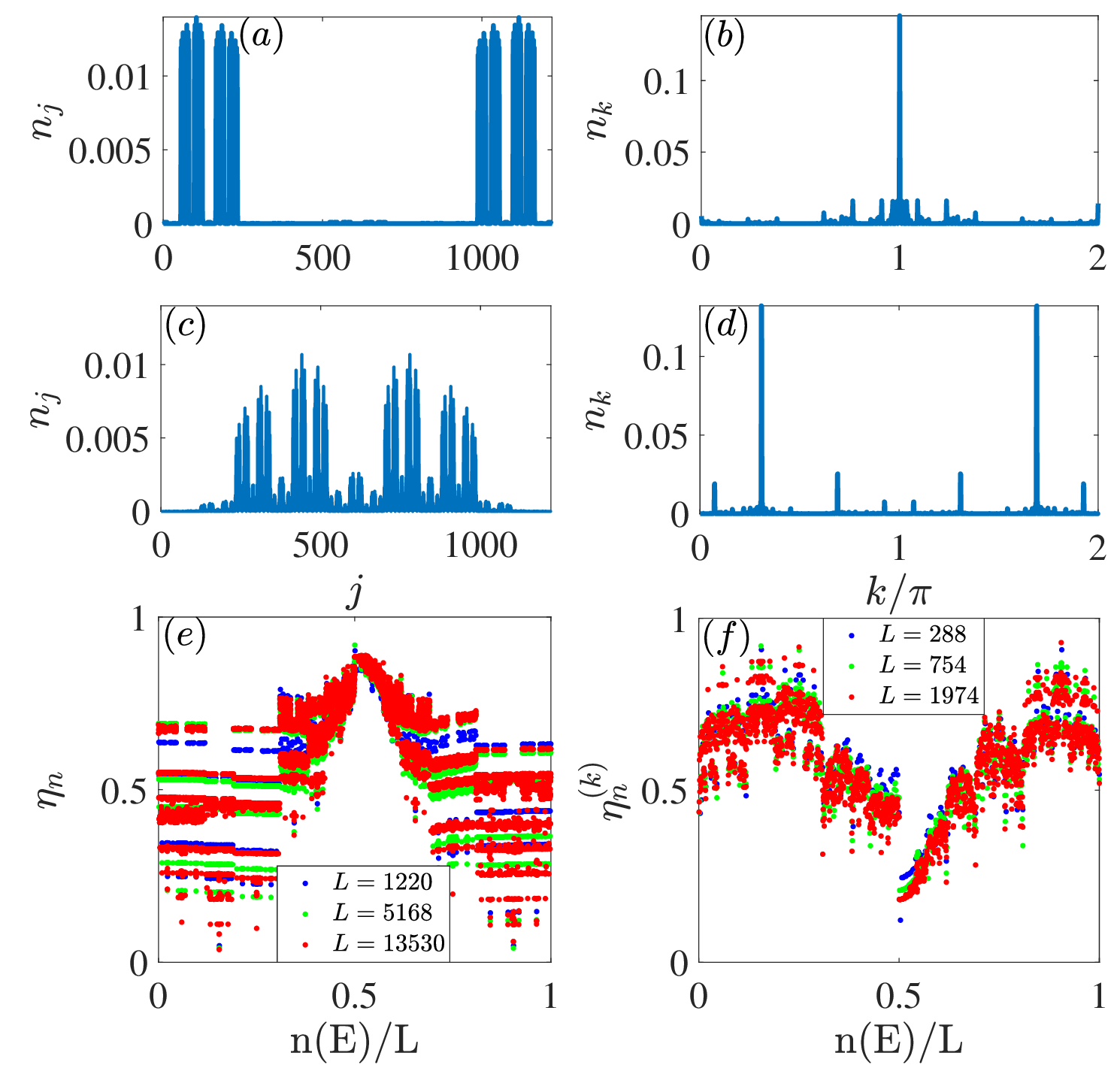}
	\end{center}
	\caption{(Color online) (a), (b) Distribution of the first with  $E\approx -3.8053$ eigenstate in the real and momentum spaces, respectively. (c), (d) Distribution of the $843$-th with  $E\approx 0.4431$ eigenstate in the real and momentum spaces, respectively. (e), (f) Fractal dimensions in the real and momentum spaces for different sizes, respectively. Here, $L=2N=1220$ for [(a)$-$(d)], $\lambda=3$, and $\beta  \to \infty$.}\label{Fig4}
\end{figure}

For a standard Fibonacci chain, it is known that all the eigenstates are multifractal for any values of $\lambda\ne 0$, which exhibit a self-similar structure. For the Hamiltonian Eq.(\ref{eq1}), when $\beta \to \infty$, the on-site potentials for the even sites reduce to two discrete values according to a Fibonacci sequence, and the amplitude of the odd sites is zero. One can easily find that in the $\beta \to \infty$ limit, the corresponding reduced characteristic Eq. (\ref{eq3}) becomes a standard Fibonacci Hamiltonian for $E\ne 0$, which means that except the state with $E=0$, all the eigenstates of the mosaic Fibonacci model are multifractal. In this subsection, we perform the numerical calculation by taking the on-site potentials of the even sites as $V_{2m}=-\mathrm{sgn}\left[ \cos{(2\pi \alpha m) - \cos{(\pi\alpha)}} \right]$ in the $\beta \to \infty$ limit, where $\mathrm{sgn}$ is the sign function. We show the fractal dimension $\eta_n$ in the real space as a function of $E_n$ and $\lambda$ for $\beta \to \infty$ in Fig. \ref{Fig3}(a). Except for the $E=0$ state, which corresponds to the $N+1$-th eigenstate, the values of fractal dimensions of all the states exhibit a multifractal characteristic. Due to the finite-size effect, one can find the red region in the small $\lambda$ shown in Fig. \ref{Fig3}(a). We believe that with the increase of system size, the red region would shrink to a point that is the state with $E=0$. Figure \ref{Fig3}(b) shows the scaling of the IPRs for different eigenstates with $\lambda=3$ in the $\beta\to \infty$ limit. The IPR of the zero-energy state (the $N+1$-th state) decreases as $1/L$ to $0$ with the increase of system size. For the other states ($n\ne N+1$), $\mathrm{IPR}^{(n)} \propto L^{-a}$ with $a\in (0,1)$. One can find that the fractal dimensions of the states in the band-center region are larger than that in the band-edge regions. The MIPRs for different $\lambda$ as the function of $1/L$ are shown in Fig. \ref{Fig3}(c). The numerical results imply that the systems with different $\lambda$ are always in the critical regime.

To further confirm the multifractal states, we contrast the wave function's distributions in real and momentum spaces with $\lambda=3$ and $L=2N=1220$ in Figs. \ref{Fig4}(a)$-$\ref{Fig4}(d). We choose the ground state in the band-edge region and the $843$-th state in the band-center region to discuss. Both states display multifractal behavior in real and momentum spaces. Figures \ref{Fig4}(e) and \ref{Fig4}(f), respectively, show the $\eta_n$ and $\eta_{n}^{(k)}$ with different $L$ and $\lambda=3$. The values are away from $0$ and $1$, except for the $E=0$ case. Our results indicate that for the $\beta \to \infty$ limit, the mosaic Fibonacci model exhibits similar localization features as the standard Fibonacci model, except for the state with $E=0$.

	\section{Cascade-like transitions by continuously controlling the knob parameter}
	
		\begin{figure}[tbp]
		\begin{center}
			\includegraphics[width=.5 \textwidth] {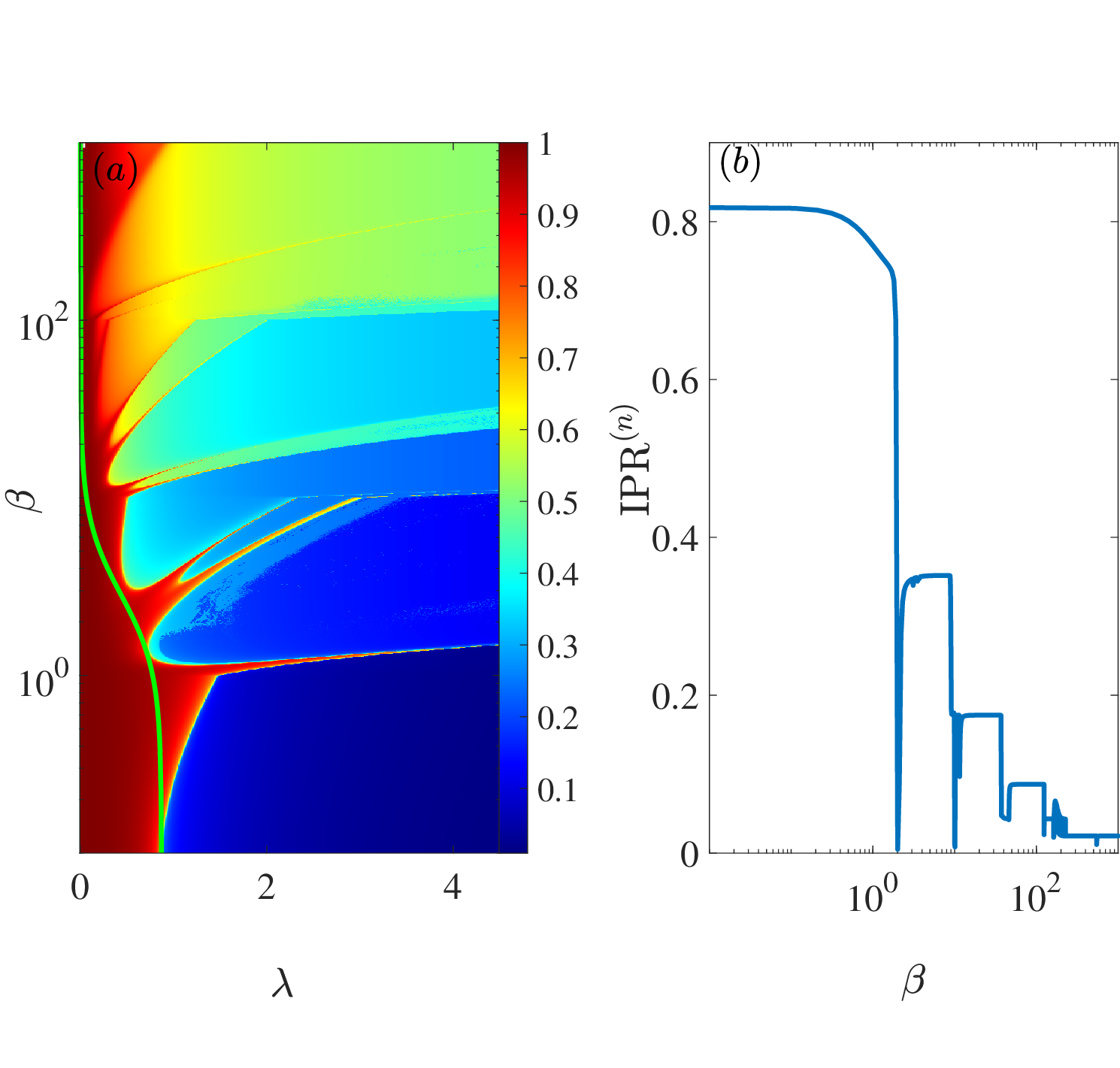}
		\end{center}
		\caption{(Color online) (a) Fractal dimension of the ground state as a function of the disorder strength $\lambda$ and the parameter $\beta$. The green line $\lambda_c$ marks the analytically obtained transition line for the ground state. (b) The IPR of the ground state as a function of $\beta$ for $\lambda=3$. Here, $L=2N=1220$.}\label{Fig5}
	\end{figure}

	\begin{figure}[tbp]
	\begin{center}
		\includegraphics[width=.5 \textwidth] {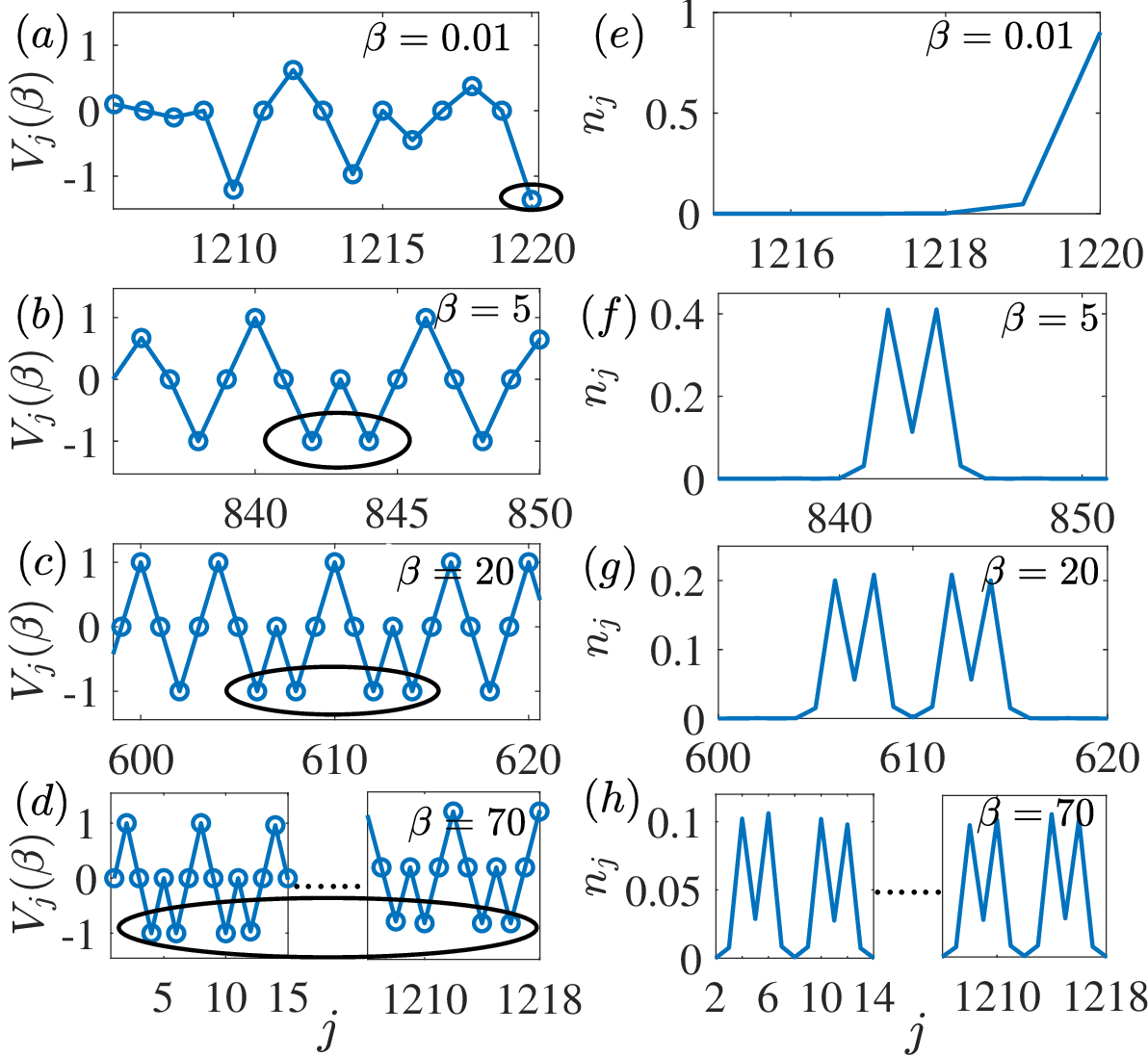}
	\end{center}
    \caption{(Color online) (a)$-$(d) Evolution of the spatial on-site potentials for different values of $\beta$. The black circles represent the minimum of on-site potentials. (e)$-$(h) Typical spatial distributions of the single-site, two-site, four-site, and eight-site localization states, respectively. Here, we choose $\lambda=3$, $L=2N=1220$, and from top to bottom  with $\beta=0.01$, $5$, $20$, and $70$, respectively.}\label{Fig6}
    \end{figure}

\begin{figure}[tbp]
	\begin{center}
		\includegraphics[width=.5 \textwidth] {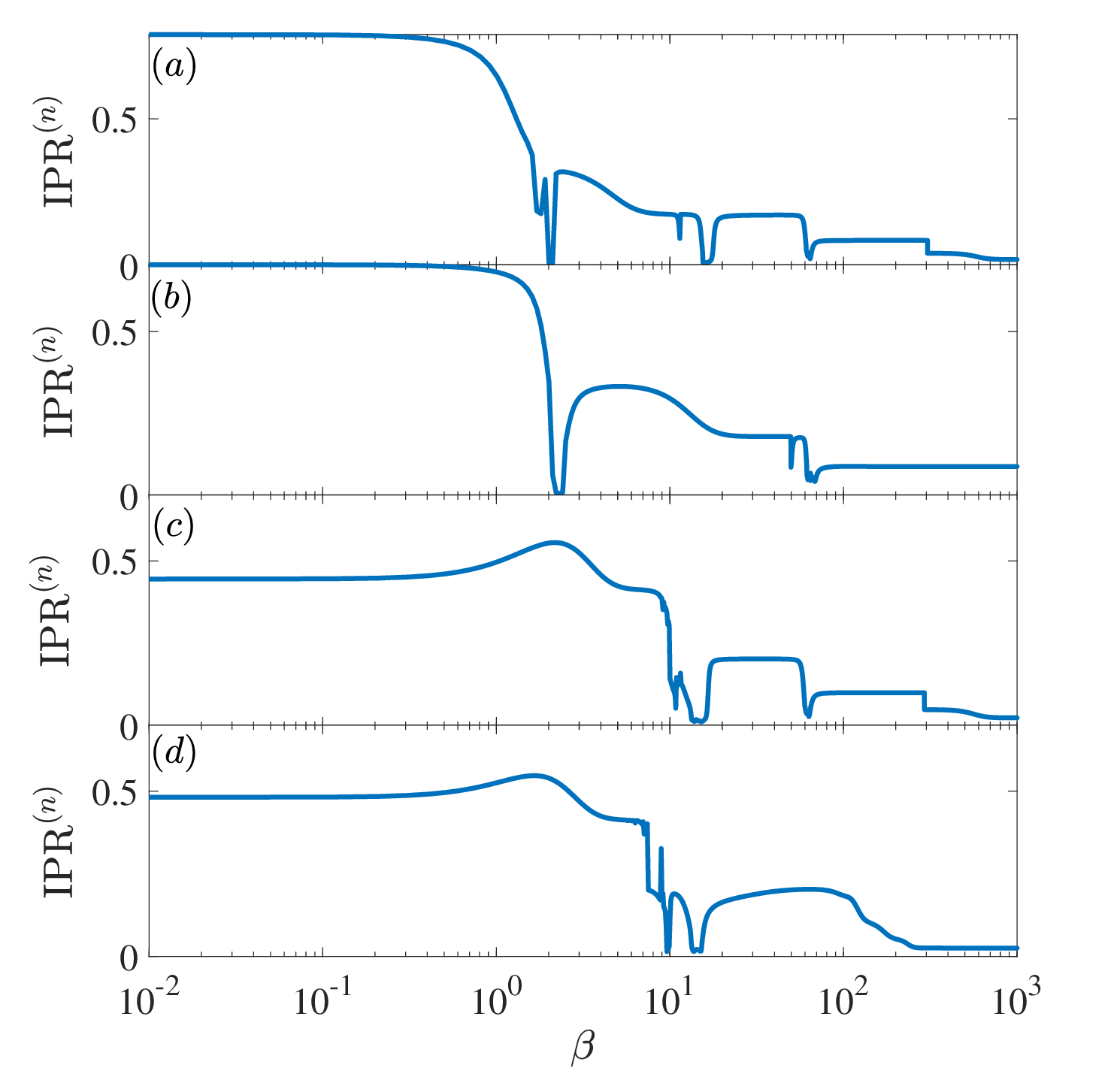}
	\end{center}
	\caption{(Color online) (a)$-$(d) The IPRs of different states localized in the band-edge regions as a function of $\beta$ for $L=2N=1220$ and $\lambda=3$ with the $89$-th, $165$-th, $283$-th, and $301$-th eigenstates, respectively.}\label{Fig7}
\end{figure}

	For a standard IAAF model, one displays the cascade of delocalization transitions from the AA limit to the Fibonacci limit with the increase of $\beta$. The phenomenon's emergence is independent of the choice of states. In the above discussions, the mosaic IAAF model of both limits exhibits distinct localization properties in the band-edge and band-center regions. The following discusses the cascade-like transitions for different states localized in different band regions with the increase of the knob parameter $\beta$ from the mosaic AA limit to the mosaic Fibonacci limit.
	
	First, we show the delocalization transition of the fractal dimension for the ground state. Figure \ref{Fig5}(a) shows the fractal dimension of the ground state as a function of the disorder strength $\lambda$ and the parameter $\beta$ with $L=1220$. According to the mosaic AA limit results, the ground state's extended-to-localized transition point is at $\lambda_{c}\approx 0.86$ for $\beta\to 0$. As seen in Fig. \ref{Fig5}(a), one can see that the extended regime gradually decreases with the increase of $\beta$. In large $\beta$ limit, the system becomes a mosaic Fibonacci model, which shows a multifractal feature for an arbitrary finite $\lambda$ except for the state with $E=0$. One can apply the generalized Avila's global theory argument in the small $\beta$ limit (see Appendix) to obtain a mobility edge $E_c$-dependent analytical result,
	\begin{equation}\label{eq8}
		\lambda_{c} = \left|\frac{2}{E_c[1-\frac{1}{6}\cos{(2\pi\alpha)}\beta^2]}\right|,
	\end{equation}
	which corresponds to the green line $\lambda_c$ shown in Fig. \ref{Fig5}(a) with $E_c\approx -2.3027 $ for the ground state. In the low $\beta$ case, this green line separates extended states from localized ones. With the increase of $\beta$, the analytical result deviates. The states in the region of the left side of the green line are extended as our numerical calculation. For large $\beta$, since the extended region is suppressed by the multifractal phase hosted by the mosaic Fibonacci limit for the ground state, the extended region greatly shrinks.
	
	As seen in Fig. \ref{Fig5}(a), for an intermediate $\lambda$, the fractal dimension does not evolve monotonously as the function of $\beta$, which displays a cascade of lobes with the lower fractal dimension values separated by the states with the maxima $\eta_n$ values. Taking $\lambda=3$ as an example, as shown in Fig. \ref{Fig5}(b), the IPR with the increase of $\beta$ displays a series of plateaux whose heights decrease in a stepwise manner, and an apparent dip emerges between two contiguous plateaux. Until the mosaic Fibonacci limit, the ground state finally evolves into a multifractal one. The phenomenon for the ground state realized by tuning the knob parameter $\beta$ from the mosaic AA limit to the mosaic Fibonacci limit corresponds to a delocalization process. The mechanism of the cascade of delocalization behaviors for the ground state is similar to that in the standard IAAF case. To better understand the mechanism of the emergence of the cascading phenomenon, we plot the on-site potentials for different $\beta$ in Figs. \ref{Fig6}(a)$-$6(d) and the corresponding ground state's density distributions in Figs. \ref{Fig6}(e)$-$6(h). Here, we take $\lambda=3$, $L=1220$, and from top to bottom $\beta=0.01$, $5$, $20$, and $70$, respectively. The black circles in Figs. \ref{Fig6}(a)$-$6(d) denote the minimum of on-site potentials. Moreover, the peaks of the ground state distributions are localized at the corresponding positions shown in Figs. \ref{Fig6}(e)$-$6(h). In the mosaic AA limit, the minimum of on-site potentials emerges at a single site [Fig. \ref{Fig6}(a)], and the corresponding ground state localized at a single site is a solid localized mode for a large $\lambda$ [Fig. \ref{Fig6}(e)], of which the IPR tends to be $1$. With the increase of the knob parameter, since a paired site potential goes down to the minimum of the potential shown in Fig. \ref{Fig6}(b), a two-site localized state turns to a new ground state [Fig. \ref{Fig6}(f)]. The region between the single-site localization and two-site localization is an extended phase corresponding to a sudden dip in the IPR. The potential values of higher-site groups sequentially become the lowest ones with $\beta$ further increasing, as seen in Figs. \ref{Fig6}(c) and 6(d), which leads to the corresponding localized states' emergence shown in Figs. \ref{Fig6}(g) and 6(h). Thus, similar cascade structures emerge with the increase of $\beta$.
	
	We also choose certain states in the band-edge regions for further discussion. As shown in Figs. \ref{Fig7}(a)$-$7(d), we show the IPRs of different states localized in the band-edge regions as the function of $\beta$ for $L=1220$ and $\lambda=3$ with the $89$th, $165$th, $283$th, and $301$th eigenstates, respectively. With the increase of the eigenvalues, the emergence of the first delocalization transition for the corresponding eigenstates is postponed. Moreover, the number of the emergence of the cascade regions is much smaller than in the ground state case. We can see the evolution of the fractal dimension of band-edge regions as the function of $\beta$ shown in Fig. \ref{Fig9} with $\lambda=3$. We observe that the lowest set of eigenenergies is squeezed into a narrow spectral window, and the delocalization is at $\beta\approx 2$. By further increasing the knob parameter, the states are localized once more. And this process repeats at the emergence of the maximum of the fractal dimension. Different bands show similar cascades at different $\beta$ for band-edge regions. It implies that the cascading transitions in the band-edge regions to the multifractal states do not happen uniformly.
	
	\begin{figure}[tbp]
		\begin{center}
			\includegraphics[width=.5 \textwidth] {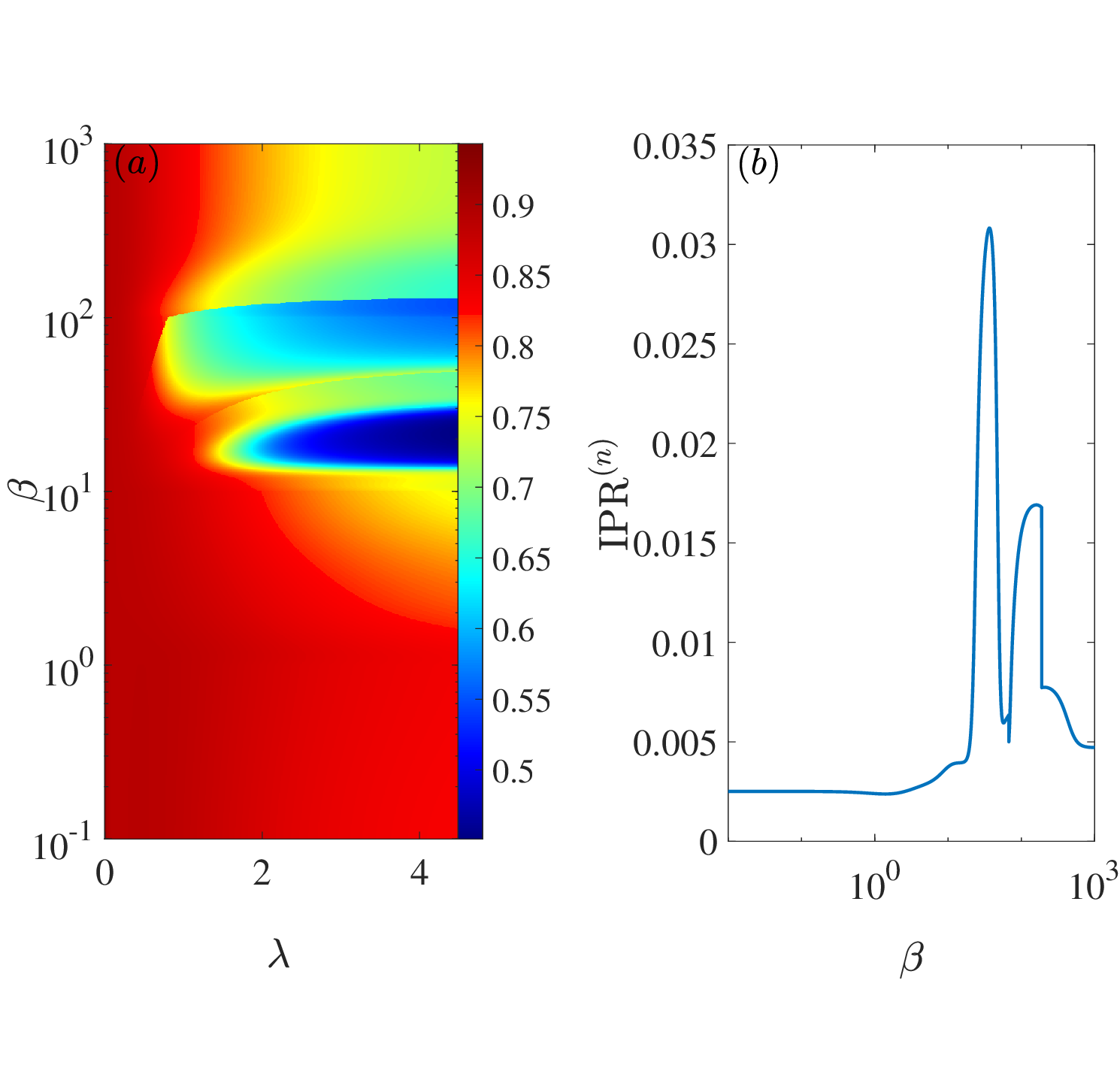}
		\end{center}
		\caption{(Color online) (a) Fractal dimension of the $843$th state as a function of the disorder strength $\lambda$ and the parameter $\beta$. (b) The IPR of the $843$th eigenstate as a function of $\beta$ for $\lambda=3$ under PBC. Here, $L=2N=1220$.}\label{Fig8}
	\end{figure}
	
	\begin{figure}[tbp]
		\begin{center}
			\includegraphics[width=.5 \textwidth] {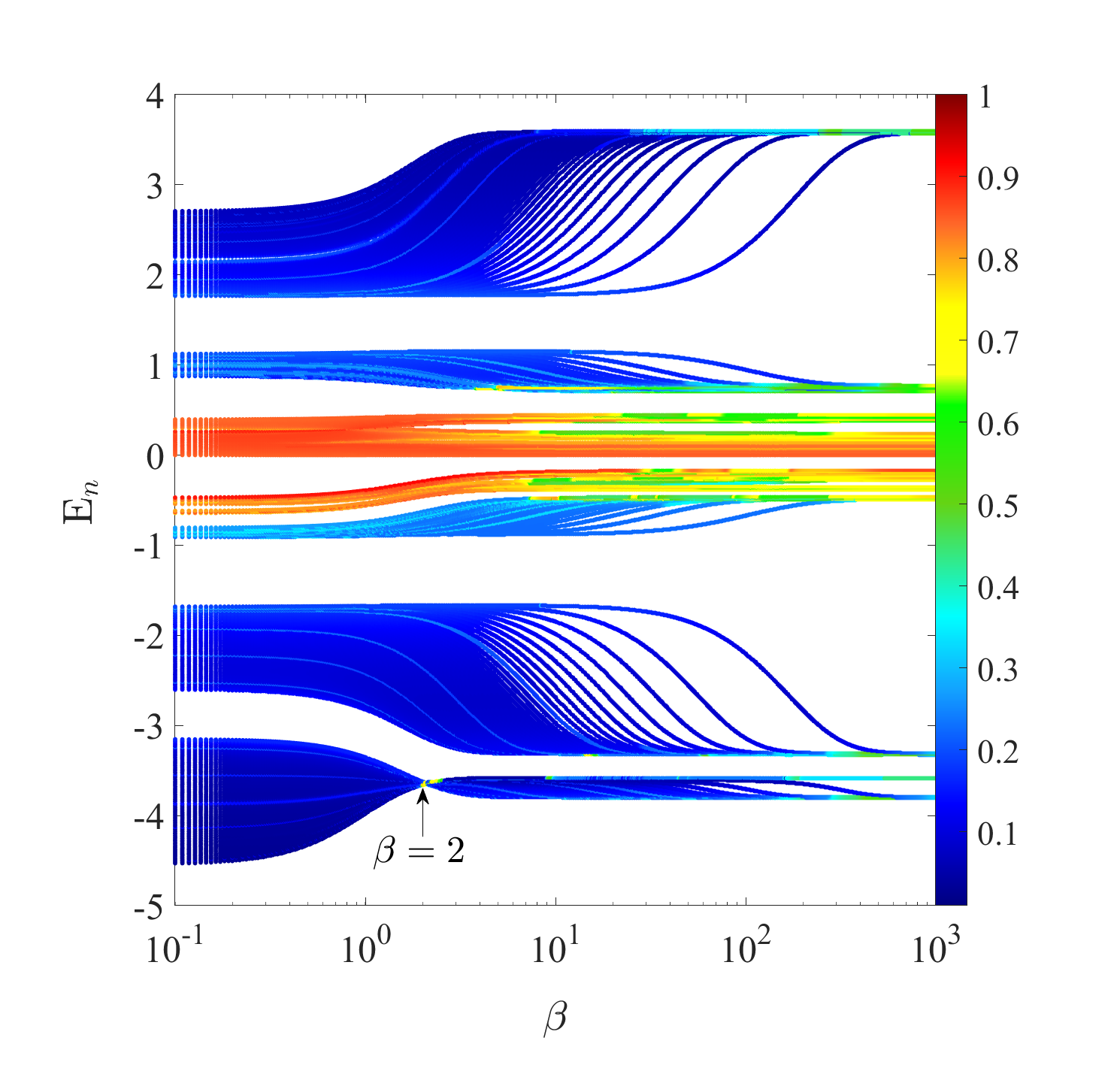}
		\end{center}
		\caption{(Color online) Fractal dimension of all eigenstates  of Eq. (\ref{eq1}) as a function of the eigenenergy and the knob parameter $\beta$. Here, $L=2N=1220$ and $\lambda=3$.}\label{Fig9}
	\end{figure}

	For the band-center region in the mosaic AA limit, the system's eigenstates are extended, while in the mosaic Fibonacci limit, all the eigenstates are multifractal except for the $E=0$ case. We expect that by tuning the knob parameter $\beta$, the states of the fractal dimension in the band-center region could exhibit a nontrivial increase. For a zero-energy state in the band-center region, according to Eq. (\ref{eq3}), the fractal dimension always keeps a unit during the increase of $\beta$. Nevertheless, as long as this zero-energy state deviates, this situation will change. We choose the $843$th state in the band-center region as an example to study its change of the fractal dimension from $\beta\to 0$ to $\beta \to \infty$ as a function of $\lambda$, which is shown in Fig. \ref{Fig8}(a) with $L=1220$. The fractal dimension displays a trivial increase with $\beta$ in the small $\lambda$ limit. However, for a large $\lambda$ case, the fractal dimension evolves non-monotonously with $\beta$ but exhibits an anomalous cascade of lobes of higher fractal dimension values separated by the regions with lower fractal dimensions. Such behavior is different from that emerges in the band-edge regions. The IPR of the $843$th state as a function of $\beta$ for the system with $\lambda=3$ and $L=1220$ is shown in Fig. \ref{Fig8}(b). In the small $\beta$, the IPR value of the $843$th state keeps small, corresponding to an extended one. With the increase of $\beta$, the value of the IPR undergoes a series of sudden increasing and decreasing processes. When $\beta$ is large enough, the IPR holds stable. As seen in Fig. \ref{Fig9}, the transition in the band-center region displays an anomalous cascade feature with the emergence of the lower fractal dimension values between the regions with higher fractal dimension values.
	
	\begin{figure}[tbp]
		\begin{center}
			\includegraphics[width=.5 \textwidth] {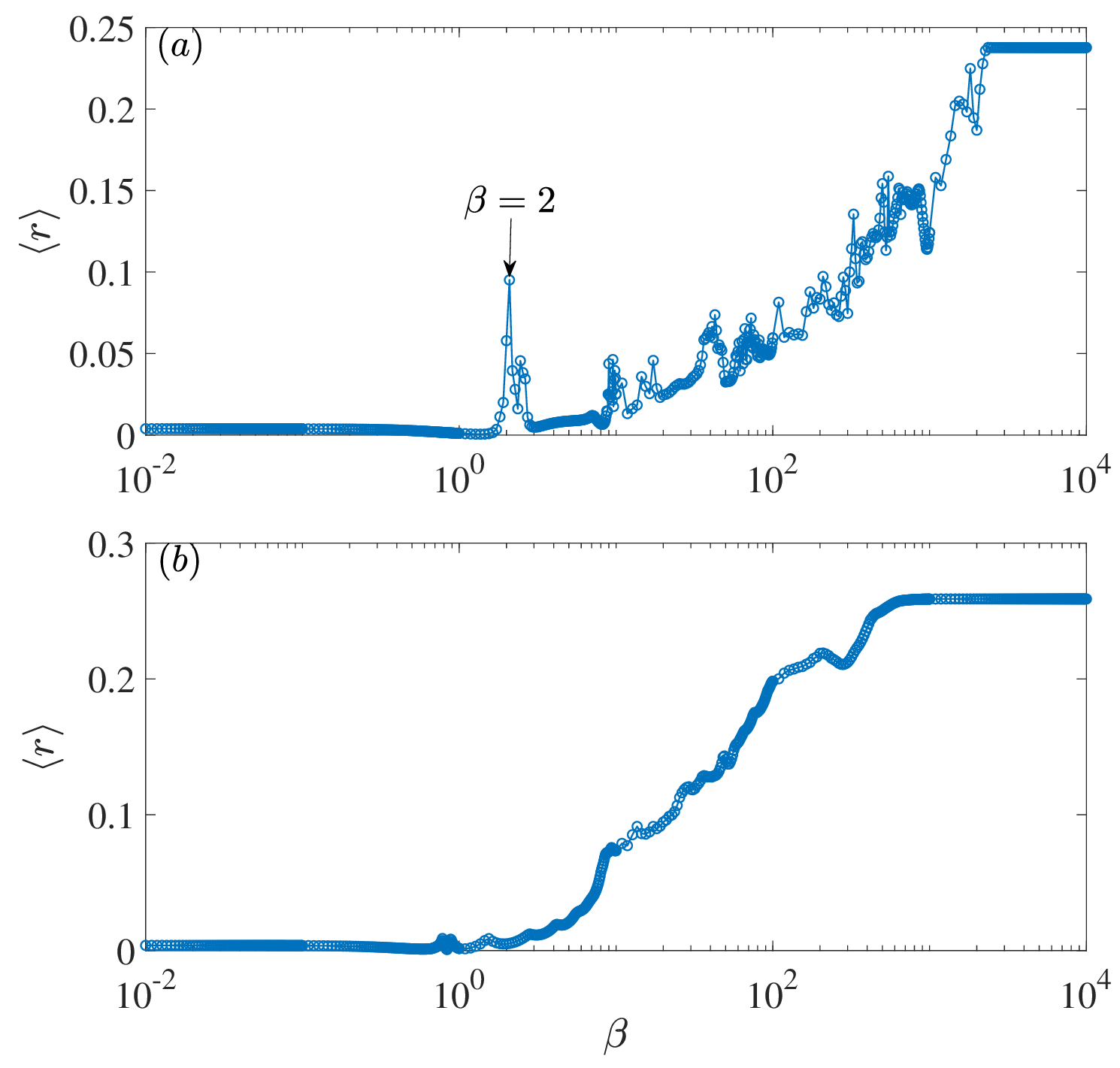}
		\end{center}
		\caption{(Color online) (a) The mean gap ratio $\langle r \rangle$ of the band-edge region as a function of $\beta$. The lowest set of eigenenergies shown in Fig. \ref{Fig9} is considered to get $\langle r \rangle$. (b) The mean gap ratio $\langle r \rangle$ of the band-center region as a function of $\beta$. We take the energy level statistics for the center of the energy band around $E=0$ with energy indexes $n\in [611,843]$. Here, $L=2N=1220$ and $\lambda=3$.}\label{Fig10}
	\end{figure}
	
	The spectral distribution of the mosaic IAAF model shows interesting properties. To obtain the spectral distribution directly, we calculate a dimensionless quantity $r_n$ \cite{Atas,Ahmed}, which is defined as
	\begin{equation}\label{eq9}
		r_n=\frac{\min(s_n,s_{n+1})}{\max(s_n,s_{n+1})},
	\end{equation}
	where $s_n=E_n-E_{n-1}$ is the spacing between the consecutive energy levels. The mean gap ratio $\langle r \rangle$ is calculated by averaging $r_n$ over different energy level regions. The mosaic AA limit has verified that most energy levels are twofold degenerate, related to the parent twofold degeneracy for $k$ and $-k$ states in the clean chain. Though the inlaid quasiperiodic potential breaks the lattice's translational symmetry, the two degeneracy is inherited in the mosaic AA limit \cite{Wang}. Due to the twofold degeneracy in the mosaic limit, the mean gap ratio $\langle r \rangle$ tends to zero in the small $\beta$ limit. Figure \ref{Fig10}(a) shows the mean gap ratio $\langle r \rangle$ of the band-edge region as the function of $\beta$ with $L=2N=1220$ and $\lambda=3$. Here, we consider the lowest set of eigenenergies shown in Fig. \ref{Fig9} to get $\langle r \rangle$. As shown in Fig. \ref{Fig10}(a), $\langle r \rangle$ approaches zero, corresponding to the twofold degeneracy feature in the band-edge region's small $\beta$ limit. Increasing $\beta$, the first peak of $\langle r \rangle$ emerges at $\beta \approx 2$. It corresponds to the emergence of the squeezed narrow spectral window in the lowest set of eigenenergies seen in Fig. \ref{Fig9}, where all the eigenstates are delocalized, and the cascading phenomena happen. Since the cascading behaviors are not uniform for different states in the band-edge regions, which has been shown in Figs. \ref{Fig7} and \ref{Fig9}, the mean gap ratio $\langle r \rangle$ displays a complex increase as the function of $\beta$, and the twofold degeneracy breaks in this region. When the system immerses into the multifractal regime in the large $\beta$ limit, one can see that the values of $\langle r \rangle$ keep stable. Figure \ref{Fig10}(b) shows $\langle r \rangle$ of the band-center region as a function of $\beta$ with $L=2N=1220$ and $\lambda=3$. Here, we take the energy level statistics for the center of the energy band around $E=0$ with energy indexes $n\in [611,843]$ for $L=1220$. In the mosaic AA limit, the twofold degeneracy leads to the values of the mean gap ratio tending to zero. When $\beta$ goes beyond $\sim 10^0$, the band-center region that we calculate begins to emerge cascading phenomena, and the degeneracy will be broken. When all the states in the band-center region shown in Fig. \ref{Fig10}(b) become multifractal in the large $\beta$ limit, the mean gap ratio $\langle r \rangle$ reaches saturation. It implies that the mosaic IAAF model's spectral distribution can help us distinguish the regimes where the cascading phenomena happen.

	In Ref. \cite{Roy}, the cascade-like transition displays a spatial modulation frequency-dependent behavior. However, are the cascading processes $\alpha$-dependent for our mosaic IAAF chain? To answer this question, we study the cascading behavior by choosing different irrational Diophantine numbers for the spatial modulation frequency $\alpha$ shown in the Appendix. For the mosaic IAAF model, the emergence of the cascadelike transitions from the mosaic AA limit to the mosaic Fibonacci limit is universal both in the band-edge and band-center regions, except for the $E=0$ case. However, the $\beta$ values where the cascading happens are the spatial modulation frequency $\alpha$-dependent.
		
	\section{Conclusion}
	
	In this paper, we study the cascade of the delocalization transitions and the emergence of the multifractal processes in a mosaic IAAF chain. In the masaic AA limit, the numerical and analytical calculations show that exact mobility edges exist and the extended states always emerge in the band-center region for an arbitrary modulation amplitude. In the mosaic Fibonacci limit, all the eigenstates of the mosaic Fibonacci model are multifractal, except for the extended state with $E=0$. Hence, there are two different ways to enter the multifractal phase, which is different from the IAAF case. By tuning $\beta$ continuously from the mosaic AA limit to the mosaic Fibonacci limit, the cascade of delocalization transition is found in the band-edge regions. With the increase of eigenvalues, the emergence of the delocalization transition for the corresponding eigenstates is postponed, and the number of the emergence of the cascade regions is much smaller than in the ground-state case. Different from the cases in the band-edge regions, an anomalous cascade feature are detected with the emergence of the lower fractal dimension values between the regions with higher fractal dimension values in the band-center region with the increase of the knob parameter $\beta$. We conclude that the cascade-like transitions of the mosaic IAAF model does not happen uniformly in different band regions.
	
	Finally, we discuss the experimental possibilities of the detection of the cascading transitions in our mosaic IAAF model. On the one hand, the IAAF model has been experimentally realized in a photonic platform. By using electron-beam lithography and dry etching to process cavity samples into quasi-1D microstructures, the cascade of delocalization transition has been observed in Ref. \cite{Goblot}. On the other hand, the mosaic models' experimental scheme has been proposed using ultracold atom technology \cite{Wang}. For our mosaic IAAF lattices, it can be realized easily based on an integrated Si${}_3$N${}_4$ photonics platform by a scanning electron microscope image of the nanophotonic device, which recently has been realized experimentally \cite{GaoJun1,GaoJun2}. One can control the width of each waveguide according to the numerical vectorial mode solver to design the desired on-site potential of each modulated site for the given knob parameter $\beta$ and modulation amplitude $\lambda$ in our mosaic IAAF system. The waveguide separation is carefully designed to keep the amplitude of the hopping term uniform. By choosing different sites of inputs and then adiabatically expanding the output array by a fan-out structure, one can realize the spatial intensity measurement of different regimes in the energy diagram. Hence, the cascaded-like processes of different band regions can be detected.
	
	\begin{acknowledgements}
		We thank X. Xia for helpful discussions. Z. Xu is supported by the NSFC (Grant No. 12375016), Fundamental Research Program of Shanxi Province (Grant No. 20210302123442), and Beijing National Laboratory for Condensed Matter Physics. This paper was also supported by NSF for Shanxi Province Grant No. 1331KSC.
		
	\end{acknowledgements}

     \section*{APPENDIX}
    
     \subsection*{1. Derivation of Eq. (8)}

	In the subsection, we apply Avila's global theory for the mosaic IAAF model by taking a small $\beta$ expansion. Avila's global theory \cite{Avila,A,You} is a theoretical framework proposed by Avila during his study on the classification of transfer matrices for Schrödinger operators. This theory has given rise to numerous mathematical conjectures and has found significant practical applications. One such application of this theory is the precise calculation of Lyapunov exponents for specific transfer matrices. For instance, the Lyapunov exponents of the AA model \cite{Shu} and the mosaic AA model \cite{Wang,Zhou} can be accurately computed using this theory. Using this theory, we can obtain the critical line $\lambda_c$, which bounds the extended phase. 
	
	First, we follow the processes of Ref. \cite{Goblot} to deal with the on-site potentials at the even sites in the small $\beta$ limit. We use Taylor unfolding to expand the potential modulation in the small $\beta$ limit and obtain the modulation potential at the even sites,
	\renewcommand{\theequation}{A\arabic{equation}}
	\setcounter{equation}{0}
	\renewcommand\thefigure{A\arabic{figure}}
	\setcounter{figure}{0}
	
	\begin{equation}\label{eqA1}
		V(x,\beta) = - \chi-\frac{1}{3}\beta^2\chi(1-\chi^2)+o(\beta^3),
	\end{equation}
	where $\chi=\cos{(2\pi\alpha x)}-\cos{(\pi \alpha)}$. Note that we use the continuous version of the on-site potential Eq. (\ref{eq2}) at the even sites defined in the main text. To return to the discrete version, we restrict the position $x$ to be a set of even numbers, {\it i.e.}, $x\to 2m$. After expanding the potentials of the even sites, one can approximate the quadratic $\beta$ term at the even sites as
	\begin{equation}\label{eqA2}
		V_{2m}(\beta) \approx  - [\chi+\frac{1}{3}\beta^2\chi U],
	\end{equation}
	where $U$ is the spatial average over a single period of the potential modulation $V(x,\beta)$ at the even sites
	\begin{eqnarray}\label{eqA3}
		U&=&\alpha\int\limits_{0}^{\alpha^{-1}}[1-\chi^2]\mathrm{d} x  \nonumber\\
		&=&\alpha\int\limits_{0}^{\alpha^{-1}}[1-(\cos{(2\pi\alpha x)}-\cos{(\pi \alpha)})^2]\mathrm{d} x \nonumber\\
		&=&-\frac{1}{2}\cos{(2\pi\alpha)}.
	\end{eqnarray}
	In this approximation, the effective potentials at the even sites remain a cosine function incommensurate with the underlying lattice, but its amplitude is altered with $\beta$. Hence, the effective Hamiltonian keeps the same shape as the mosaic AA model, but with a $\beta$-dependent modulation amplitude, {\it i.e.}, $\lambda \to \tilde{\lambda}_{\beta}=\lambda [1-\frac{1}{6}\cos{(2\pi\alpha)}\beta^2]$.
	
	Next, we apply Avila's global theory to calculate the Lyapunov exponents of the effective Hamiltonian, which has the same form as the mosaic AA model \cite{Wang}. The Lyapunov exponent $\mathcal{L}$ can be computed as
	\begin{equation}\label{eqA4}
		\mathcal{L}(E) = \lim_{N\to\infty} \frac{1}{2N} \ln{||T(E)||},
	\end{equation}
	where $N$ is the number of the quasicell, $||T(E)||$ denotes the norm of the total transfer matrix $T(E)$ for a given $E$, and $T(E)=T_{N}(E)T_{N-1}(E)\cdots T_1(E)$ with $T_m(E)$ being the local transfer matrix in one quasicell. The local transfer matrix is given by
	\begin{equation}\label{eqA5}
		T_{m}(E)=\begin{pmatrix} E+\tilde{\lambda}_{\beta} \chi & -1 \\ 1 & 0 \end{pmatrix} \times \begin{pmatrix} E & -1 \\ 1 & 0 \end{pmatrix}.
	\end{equation}
	According to Avila's global theory, we introduce a complex phase $i\epsilon$ into $\cos(2\pi\alpha m)$. The complexification of the phase is important for us since our computation relies on Avila's global theory of one-frequency analytical $SL\rm{(2,\mathbb{R})}$ cocycle \cite{Avila}. That is,
	\begin{eqnarray}\label{eqA6}
		\cos(2\pi\alpha m+i\epsilon)=\frac{e^{i2\pi\alpha}e^{-\epsilon}+e^{-i2\pi\alpha m}e^{\epsilon}}{2}.
	\end{eqnarray}
	Let $\epsilon\to +\infty$, then direct computation yields
	\begin{equation}\label{eqA7}
		T_{m}(E+i\epsilon)=\frac{\tilde{\lambda}_{\beta}}{2} e^{-i2\pi\alpha m}e^{\epsilon}\begin{pmatrix} E & -1 \\ 0 & 0 \end{pmatrix} + o(1).
	\end{equation}
	The total transfer matrix of the whole chain is
	\begin{eqnarray}\label{eqA8}
		T(E,\epsilon)&=&T_{N}(E,\epsilon)T_{N-1}(E,\epsilon) \cdots T_{1}(E,\epsilon)\notag\\
		&=&\tilde{\lambda}_{\beta}^N \prod_{m=1}^{N} \frac{e^{-i2\pi\alpha m}}{2} (e^{\epsilon})^N\begin{pmatrix} E & -1 \\ 0 & 0 \end{pmatrix}^{N}
	\end{eqnarray}
	Thus, the norm of the total transfer matrix is
	\begin{equation}\label{eqA9}
		||T(E,\epsilon)||= \lvert \frac{1}{2} \tilde{\lambda}_{\beta} e^{\epsilon}E \rvert^N.
	\end{equation}
	Combining Eqs. (\ref{eqA4}) and (\ref{eqA9}), we can obtain
	\begin{eqnarray}\label{eqA10}
		\mathcal{L}_{\epsilon}(E) = \frac{1}{2} \left(\ln\lvert \mu \rvert +\epsilon \right),
	\end{eqnarray}
	with $\mu=\tilde{\lambda}_{\beta}E/2$. Avila's global theory \cite{Avila,A,You} shows that, as a function of $\epsilon$, $2\mathcal{L}_{\epsilon}(E)$ is a convex, piecewise linear function, and their slopes are integers, which implies $2\mathcal{L}_{\epsilon}(E)=\max{\{\ln\lvert \mu \rvert+\epsilon,2\mathcal{L}_{0}(E)\}}$. Moreover, Avila's global theory tells us that, $E$ does not belong to the spectrum of the Hamiltonian, if and only if $\mathcal{L}_0(E)\textgreater 0$, and $\mathcal{L}_{\epsilon}(E)$ is an affine function in the neighborhood of $\epsilon=0$. Consequently, when $E$ is localized in the spectrum, we have
	\begin{equation}\label{eqA11}
		2\mathcal{L}_{0}(E)=\max\Big\{\ln\lvert \mu \rvert,0\Big\}.
	\end{equation}
	For a given eigenvalue $E$, when $\lvert \mu \rvert > 1$, the localization length
	\begin{equation}\label{eqA12}
		\zeta(E)=\frac{1}{\mathcal{L}_0(E)}=\frac{2}{\ln\lvert \mu\rvert }
	\end{equation}
	is a finite value, which denotes that the corresponding state is localized. When $\lvert \mu\rvert <1$, $\zeta\to\infty$ corresponds to a delocalized one. Thus, by $\lvert \mu \rvert= 1$, one can obtain the critical line
	\begin{equation}\label{eqA13}
		\lambda_c=\left| \frac{2}{E_c (1-\frac{1}{6}\cos(2\pi\alpha)\beta^2)}\right|.
	\end{equation}
	Note that this result is suited for the small $\beta$ limit. For a large $\beta$, the result deviates.
    
    \subsection*{2. Effects of the spatial modulation frequency on cascading transitions}
    
    \begin{figure}[tbp]
    	\begin{center}
    		\includegraphics[width=.5 \textwidth] {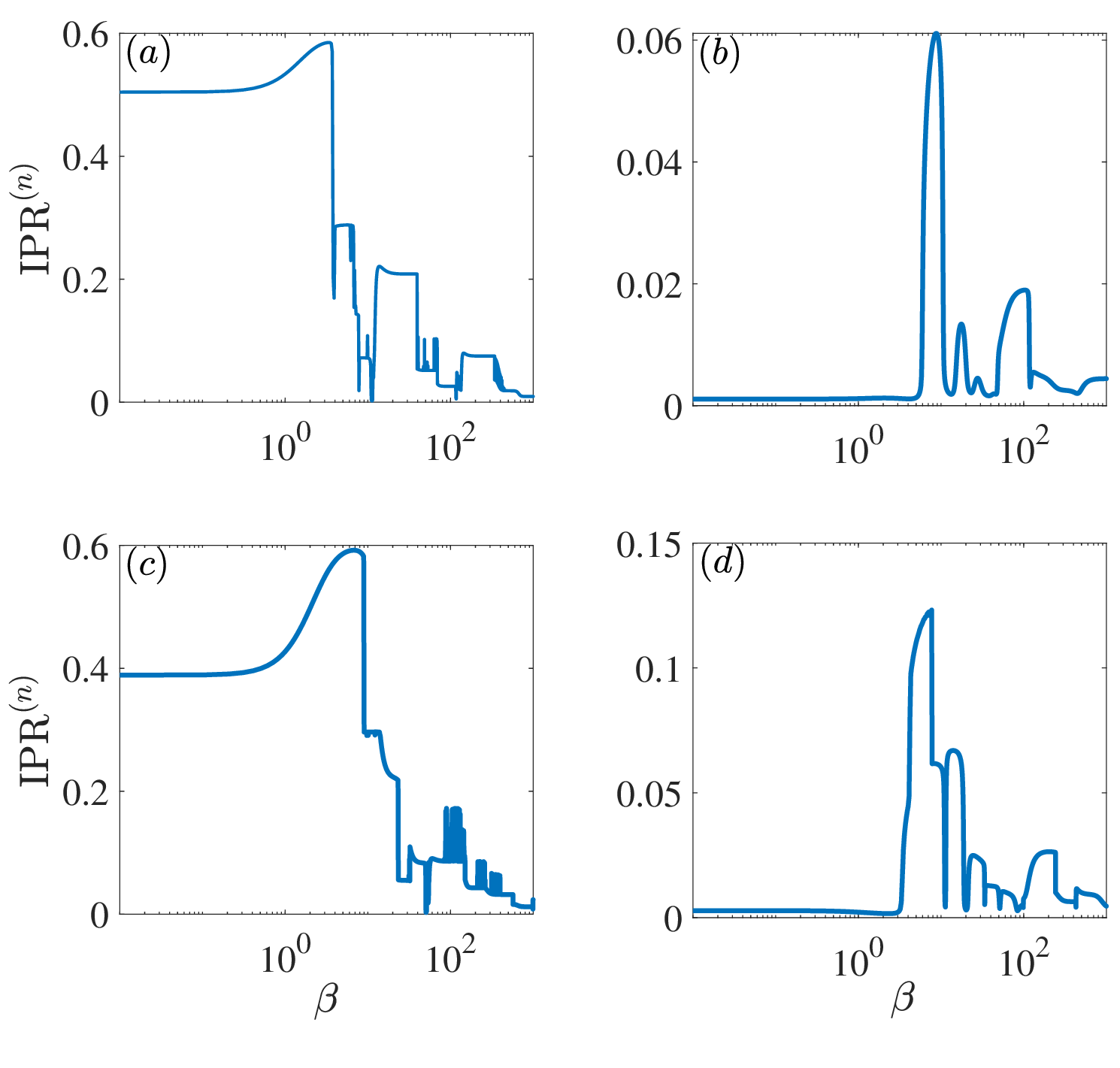}
    	\end{center}
    	\caption{(Color online) (a), (b) The IPRs of the ground state and the $1155$th state as a function of $\beta$ for $\lambda=2$ and $L=1970$ with $\alpha_s$, respectively. (c), (d) The IPRs of the ground state and the $829$th state as a function of $\beta$ for $\lambda=2$ and $L=2378$ with $\alpha_b$, respectively.}\label{Fig11}
    \end{figure}

    To study the effects of the spatial modulation frequency $\alpha$ on cascading transitions, we choose other irrational Diophantine numbers for our discussion, which are different from the choice in the main text, with $\alpha$ being the inverse of the golden mean $(\sqrt{5}-1)/2$. To obtain the metallic mean family of the irrational Diophantine number, we consider a generalized $\kappa$-Fibonacci sequence, $F_{v+1}=\kappa F_v+F_{v-1}$, with $F_0=0$ and $F_1=1$. The irrational number $\alpha$ controlling the spatial modulation frequency can be obtained by the limit $\alpha = \lim_{v\to\infty} F_{v-1}/F_{v}$ with $\kappa=1,2,3,\cdots$, which can yield the metallic mean family. For $\kappa=1$, we can obtain the golden mean $\alpha_g = (\sqrt{5}-1)/2$, which is studied in the main text. When $\kappa=2$ and $3$, one can get the silver mean $\alpha_s=\sqrt{2}-1$ and the bronze mean $\alpha_b=(\sqrt{13}-3)/2$, respectively. We take the system size $L=2F_v$ and the rational approximation $\alpha=F_{v-1}/F_v$ under PBCs. In this part, we consider the cascading features for different choices of $\alpha$ in the band-edge and band-center regions, respectively.

    As a concrete example, in the band-edge region, we choose the ground state' IPRs as a function of $\beta$ with $\lambda=2$ shown in Fig. \ref{Fig11}(a) for $\alpha_s$ and Fig. \ref{Fig11}(c) for $\alpha_b$, respectively. Compared with the $\alpha_g$'s results in the main text, we can see the emergence of the cascading phenomena is universal, but the values of $\beta$ where the cascading happens depend on the choice of $\alpha$. In the band-center region, we choose the $1155$th eigenstate's IPR for $\alpha_s$ and $L=1970$ and the $829$th eigenstate's IPR for $\alpha_b$ and $L=2378$ as a function of $\beta$ with $\lambda=2$ shown in Figs. \ref{Fig11}(b) and \ref{Fig11}(d), respectively. In the band-center region, the cascading processes from the mosaic AA limit to the mosaic Fibonacci limit also display a $\alpha$-dependent behavior. According to our numerical results, for the mosaic IAAF model, the emergence of the cascade-like transitions from the mosaic AA limit to the mosaic Fibonacci limit is universal both in the band-edge and band-center regions. However, the $\beta$ values where the cascading happens are the spatial modulation frequency $\alpha$ dependent.

\end{document}